# Emergency DC Power Support Strategy Based on Coordinated Droop Control in Multi-Infeed HVDC System

Ye Liu, Chen Shen*, Yankan Song, Jun Yan
Department of Electrical Engineering
Tsinghua University
Beijing, China
liuye_thu@foxmail.com

*Abstract*—With the complex hybrid AC-DC power system in China coming into being, the HVDC faults, such as DC block faults, have an enormous effect on the frequency stability of the AC side. In multi-infeed HVDC (MIDC) system, to improve the frequency stability of the receiving-end system, this paper proposes an emergency DC power support (EDCPS) strategy, which is based on a designed droop characteristic between the active power of LCC-HVDC lines and the receiving-end system frequency. Then, a coordinated optimization method is proposed to determine the droop coefficients of the MIDC lines. To test the proposed EDCPS method, the electromagnetic transient (EMT) model of a MIDC system is built on the CloudPSS platform, and the test results verify the effectiveness of the proposed EDCPS strategy.

*Index Terms*--Emergency DC power support, P-f droop control, multi-infeed LCC-HVDC system, frequency control.

## I. INTRODUCTION

With the development of multi-type HVDC transmission technologies, the evolutionary trend of power grid is a large-scale complex hybrid AC-DC system [1]. Multi-infeed HVDC (MIDC) systems widely exist in hybrid AC-DC system due to the considerable power demand of the receiving-end system. However, in the MIDC system, various HVDC faults, such as the block faults, will cause the active power loss in the receiving-end system, which poses a serious threat to the frequency stability of the AC side [2].

The traditional approach to solve the aforementioned problem is load shedding. However, the power loss caused by the HVDC block faults is considerable because of the large transmission capacity of HVDC, and therefore a large amount of load will be shed to maintain the frequency stability. Furthermore, the line commutated converter based HVDC (LCC-HVDC) system in MIDC system can achieve fast regulation of active power, and therefore the emergency DC power support (EDCPS) strategy based on LCC-HVDCs can be considered to alleviate the frequency stability problems [3], [4].

The existing EDCPS strategies are executed mainly by means of centralized decision and control. In [5], an adaptive dynamic surface control based EDCPS strategy is proposed and its small signal stability is proved. In [6], an emergency frequency control strategy considering LCC-HVDC and centre of inertia is proposed, and the adaptive backstepping control is utilized to guarantee its robustness. However, the above two methods are designed for AC-DC parallel interconnected system, which are only applicable for two synchronous subsystems. In [7], for the Ha-Zheng HVDC system and the receiving-end Central China grid, a response-based AC-DC coordinated control strategy is proposed, which combines EDCPS with the load shedding. A control strategy for the ORMOC-NAGA HVDC system to improve the AC frequency stability is proposed in [8]. However, the methods in [7] and [8] are based on centralized control which requires control centers, and this method cannot ensure a rapid support to the frequency stability in the case of communication delay or communication failure. In summary, current researches have several drawbacks as follows:

- The EDCPS strategy is not considered especially for the MIDC system where the sending-end systems and the receiving-end system are asynchronous.
- The EDCPS strategy with more efficiency and less communication dependence requires a decentralized control strategy for every LCC-HVDC.

In order to solve the above drawbacks, this paper proposes a novel EDCPS strategy for MIDC system based on droop control, which is a decentralized control strategy widely applied in VSC-MTDC systems and microgrids [9-11]. A P-f droop characteristic between the active power of LCC-HVDC lines and the frequency of receiving-end system is designed and applied in MIDC system. Then, a coordinated optimization method is proposed to determine the optimal droop coefficients of the EDCPS strategy.

The rest of the paper is organized as follows. Section II introduces the droop control design of LCC-HVDC systems.

This work is supported by National Key Research and Development Program of China (2016YFB0900600) and Technology Projects of State Grid Corporation of China (52094017000W).

Section III introduces the proposed novel EDCPS strategy in a MIDC system and the coordinated optimization method. Then, the P-f droop control and the EDCPS strategy are tested in Section IV based on the CloudPSS platform. Section V provides the conclusion.

## II. Droop Control of LCC-HVDC

### A. Introduction to LCC-HVDC System

The LCC-HVDC system is widely applied in China's power grid because of its large transmission capacity and long transmission distance. Fig. 1 shows the steady-state model of a LCC-HVDC system. the AC bus is connected to the 6-pulse bridge converter via the corresponding converter transformer, and the rectifier is connected to the inverter via the DC line.

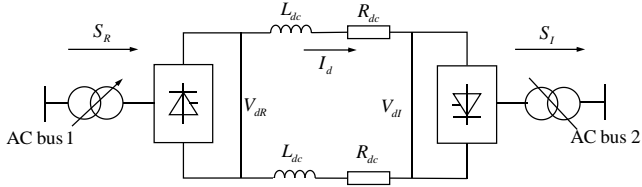

Figure 1. The steady-state model of a LCC-HVDC system.

Considering the control system of the LCC-HVDC system, the classical control mode is that the rectifier adopts the constant current control while the inverter adopts the constant γ angle control. Besides, the constant power control could be added to the rectifier to regulate the transmission power. There exists:

$$I_{order} = \frac{P_{order}}{U_d} \tag{1}$$

where $P_{order}$ is the active power set-point of the LCC-HVDC line, and $I_{order}$ is the current set-point.

As shown in Fig. 1, the steady-state equations of the LCC-HVDC are given in (2).

$$\begin{cases} V_{dR} = V_{d0R}\cos\alpha - \frac{3}{\pi}N_b X_{cR} I_d \\ V_{dI} = V_{d0I}\cos\gamma - \frac{3}{\pi}N_b X_{cI} I_d \\ V_{dR} = V_{dI} + R_{dc} I_d \\ P_{dR} = N_P V_{dR} I_d \\ P_{dI} = N_P V_{dI} I_d \end{cases} \tag{2}$$

where $V_{dR}$ and $V_{dI}$ are the DC voltages of the rectifier and the inverter respectively, $I_d$ is the DC current, $P_{dR}$ and $P_{dI}$ are the DC active powers of the rectifier and the inverter respectively, $N_p$ is the pole number, and $N_b$ is the bridge number for each pole.

The DC current and DC voltage can be regulated fast according to the reference values because of the LCC-HVDC control characteristics. Therefore, the transmission power of LCC-HVDC can also achieve fast regulation according to (2), which founds the technical basis for the EDCPS strategy.

### B. Design of the P-f Droop Control for LCC-HVDC

Generally, the droop characteristic describes the relevance between two variables. Therefore, we can explore the droop characteristic existing in LCC-HVDC system through the static operation characteristic.

The static operation characteristic of LCC-HVDC is shown in Fig. 2. The normal operation curves of the LCC-HVDC are shown as I and II in Fig. 2, where the vertical parts of the curves represent constant current control, and the rest parts are constant α control and constant γ control respectively. The intersection point A indicates that the rectifier of the HVDC system adopts constant current control, while the inverter adopts constant γ control during the normal operation.

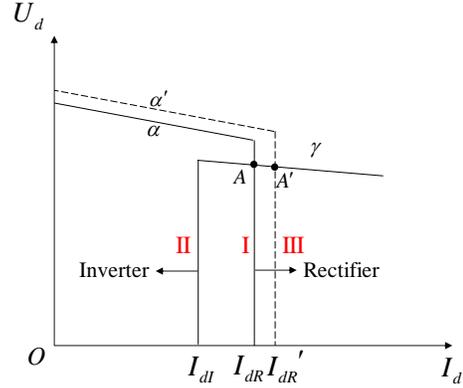

Figure 2. The static operation characteristic of LCC-HVDC.

Considering a MIDC system, in the presence of the private communication network specifically built for control and protection, the frequency signal of the receiving-end system can be quickly transmitted to the pole control station in the sending-end system, and then the active power of LCC-HVDC can be regulated by the designed active power-frequency (P-f) droop control for EDCPS when serious faults occur.

The regulation of the active power of LCC-HVDC is realized by changing the power set-point $P_{order}$ of the pole control layer, which is simply represented as:

$$\dot{P}_{DC} = \frac{1}{T_{DC}}(-P_{DC} + P_{order}) \tag{3}$$

When the $P_{order}$ increases, the $I_{order}$ will also increase assuming that $U_d$ is constant according to (1), and then the PI controller will cause a decrease in α angle to $\alpha'$, which make the static operation point A changes to $A'$, and the curve I changes to III. Eventually, the active power of LCC-HVDC increases as the $P_{order}$ increases. In addition, because of the constant γ control in inverter, the DC voltage will drop slightly according to (4).

$$V_{dI} = \frac{3\sqrt{2}}{\pi}N_b K_{TI} V_{aI}\cos\gamma - \frac{3}{\pi}N_b X_{cI} I_d \tag{4}$$

In summary, the designed droop characteristic of LCC-HVDC can be represented as the $P_{order}$-f (P-f for short) droop, as shown in Fig. 3, where $\omega = 2\pi f$. The P-f droop control equation of LCC-HVDC can be represented by (5).

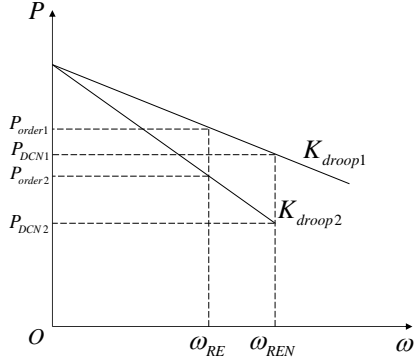

Figure 3.  The P-f droop characteristic of LCC-HVDC.

$$P_{order} = P_{DCN} + K_{droop}(\omega_{REN} - \omega_{RE}) \quad (5)$$

where $K_{droop}$ is the droop coefficient of the LCC-HVDC system, $\omega_{RE}$ and $\omega_{REN}$ are the angular frequency and its rated value of the receiving-end system respectively. Add the designed droop control to the HVDC control system, and the hierarchical control framework can be obtained, as shown in Fig. 4.

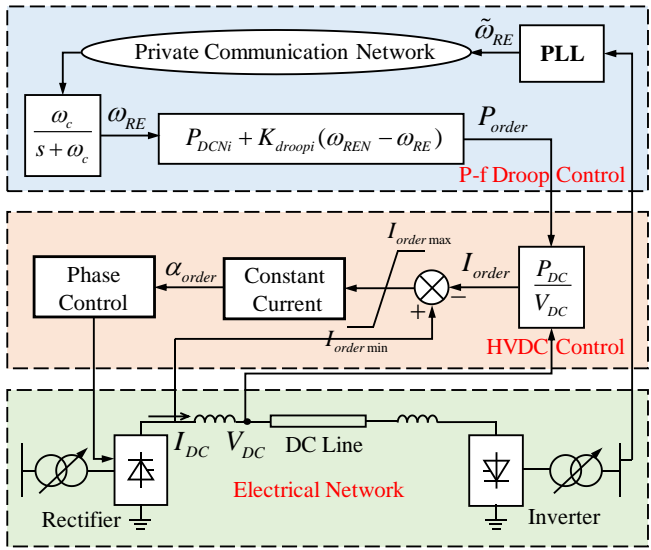

Figure 4.  The hierarchical control framework of LCC-HVDC.

The droop control does not work when the system operates in normal state. When a frequency stability problem occurs in the receiving-end system, the *P-f* droop control takes effect and the active power of LCC-HVDC will increase. In the control framework, the private communication network plays an important role which passes the frequency signal of the receiving-end system to the rectifier.

## III.  THE EDCPS STRATEGY IN MIDC SYSTEM

### A.  Procedure of the Proposed EDCPS Strategy

Based on the *p-f* droop control of the LCC-HVDC system, the procedure of developing the proposed EDCPS strategy in a MIDC system is shown in Fig. 5. And the EDCPS strategy can be executed in the following steps.

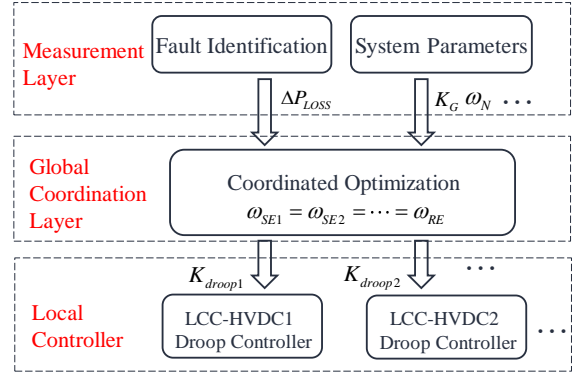

Figure 5.  The development procedure of the EDCPS strategy.

- **Droop coefficients initialization**. When the system operates steadily, set the initial values of droop coefficients for each decentralized local controller, which is located at the pole control station of LCC-HVDC system.

- **Fault identification**. Once the fault is identified successfully through the feature extraction [12], the fault information, such as the power loss, will be sent to the global coordination layer, which is located at the dispatch center of the MIDC system.

- **Coordinated optimization**. To achieve the best performance of EDCPS strategy, droop coefficients need to be optimized. And the optimization problem will be explain in the following section.

- **Droop coefficients updating**. Transmit the optimized droop coefficients to local droop controllers of LCC-HVDC lines for a better power support to the receiving-end system.

### B.  Coordinated Optimization for Droop Coefficients

Generally, a MIDC system containing *n* LCC-HVDCs and *n* sending-end systems is shown in Fig. 6.

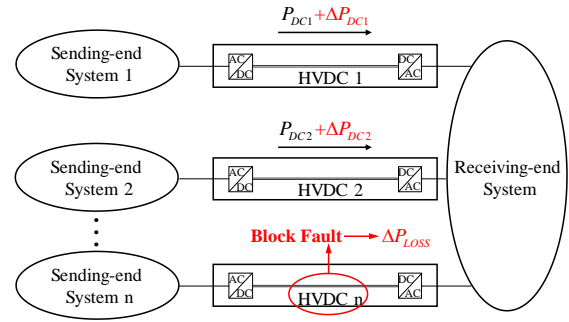

Figure 6.  The topology of a MIDC system with n LCC-HVDC lines.

Supposing that the block fault occurs to the n-th HVDC line, and the power loss of receiving-end system is $\Delta P_{LOSS}$, then a coordinated optimization is executed to determine the droop coefficients of other (n-1) LCC-HVDC lines. In order to balance the power loss caused by the block fault among the receiving-end and sending-end systems, the frequencies of receiving-end and sending-end systems are supposed to be equal after the EDCPS transient process. Besides, the load shedding power should be minimized to take full advantage of the power regulation characteristic of the LCC-HVDC system. Therefore, the objective function is:

$$\min \ (\omega_{SE1}-\omega_{SE2})^2+\cdots+(\omega_{SEn-2}-\omega_{SEn-1})^2 \\ +(\omega_{SEn-1}-\omega_{RE})^2+M\cdot\Delta P_L \quad (6)$$

where $\omega_{SE}$ is the angular frequency of the sending-end system, $\Delta P_L$ is the load shedding power, and M is a big number as the penalty coefficient. The constraints are as follows:

$$s.t. \quad \underline{\omega}_{SEi} \leq \omega_{SEi} \leq \overline{\omega}_{SEi} \quad (a) \\
\underline{\omega}_{RE} \leq \omega_{RE} \leq \overline{\omega}_{RE} \quad (b) \\
K_{GSEi}(\omega_{SENi}-\omega_{SEi}) = \Delta P_{DCSEi} \quad (c) \\
K_{GRE}(\omega_{REN}-\omega_{RE}) = \Delta P_{LOSS} - \sum_{i=1}^{n-1} \Delta P_{DCi} - \Delta P_L \quad (d) \\
P_{DCi} + \Delta P_{DCi} \leq k_{max} P_{DCNi} \quad (e) \\
\Delta P_{DCi} = K_{droopi}(\omega_{REN}-\omega_{RE}) \quad (f) \\
\Delta P_{LOSS} \geq \sum_{i=1}^{n-1} \Delta P_{DCi} + \Delta P_L \quad (g) \\
(i=1,2,\cdots,n-1)$$

where (7)(a)-(7)(d) are the frequency constraints of subsystems, (7)(e) is the maximum transmission power constraints of HVDC lines, and (7)(f) is the droop control equation. With the solution of the optimization problem, the optimal droop coefficients are provided for the EDCPS strategy.

## IV. CASE STUDY

### A. Scenario 1: Droop Characteristic of LCC-HVDC

To verify the P-f droop characteristic of LCC-HVDC system, the EMT model of the HVDC system, as shown in Fig. 1, is built on the CloudPSS platform, which is a monopolar 12-pulse LCC-HVDC model [13]. The initial operation point of LCC-HVDC is:

$U_{DC}$=600kV; $I_{DC}$=1.1kA; $P_{DC}$=660MW; $P_{order}$=0.66

Since the designed droop control is based on the relevance between variables, this section mainly verifies whether the trend of each variable is consistent with the theoretical analysis after the system parameters change. The droop coefficient of the P-f droop control is set to 20 (p.u.). Superpose a step response from 50Hz to 49.65Hz to the frequency signal, and the active power of LCC-HVDC is shown in Fig. 7.

As shown in Fig. 7, the LCC-HVDC system completes initialization around 1s. At 4s, the step response of frequency signal causes rapid increase in active power, which illustrates the feasibility of EDCPS strategy and the effectiveness of the droop characteristic. The voltage and current of HVDC and the α angle of the rectifier are shown in Fig. 8.

With the increase of active power, the DC current increases, the DC voltage decrease slightly, and the α angle decrease during the transient process, which is consistent with the theoretical analysis in section II.

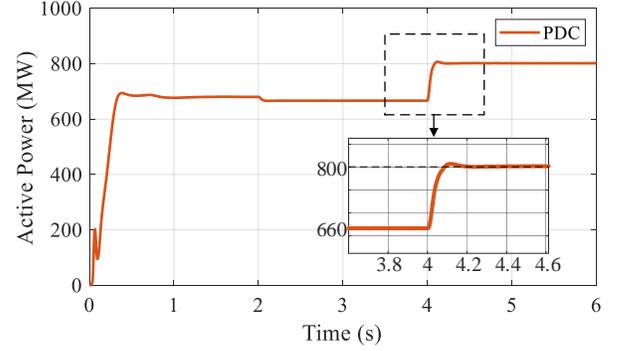

Figure 7.  The active power of the LCC-HVDC system.

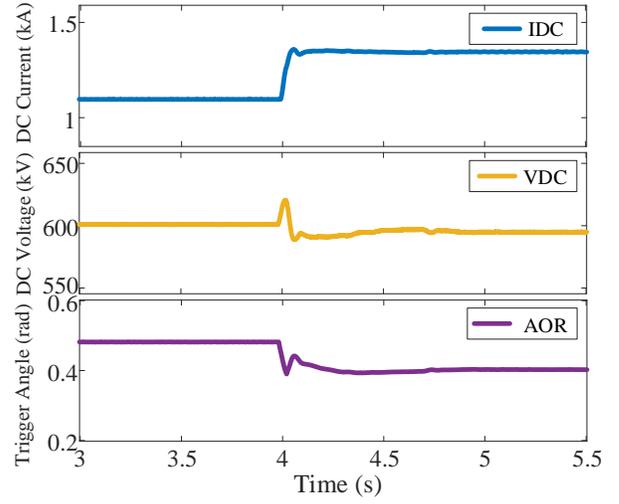

Figure 8.  The electrical variables of the LCC-HVDC system.

### B. Scenario 2: EDCPS strategy in MIDC System

The EMT model of a MIDC system containing 4 LCC-HVDC lines is built on the CloudPSS platform, as shown in Fig. 6, in which the sending-end systems are equivalent to a single generator as the centre of inertia (COI) model, and the receiving-end system is the IEEE 39-bus system. The active power of each LCC-HVDC line is set as: $P_{DC1}$ = 660MW, $P_{DC2}$ = 630MW, $P_{DC3}$ = 650MW, $P_{DC4}$ = 540MW.

Set a block fault happen to HVDC4 at about 8s, and therefore $\Delta P_{LOSS}$ = 540MW. Once the fault is identified successfully, the optimal droop coefficients obtained by solving the optimization (6) and (7) are:

$K_{droop1}$ = 30, $K_{droop2}$ = 25, $K_{droop3}$ = 29 (p.u.)

After updating the droop coefficients, the EDCPS strategy is executed to ensure the frequency stability. The frequencies of receiving-end system in three subcases are shown in Fig. 9, where 1) there is no droop control in subcase 1, 2) only HVDCs have droop control in subcase 2, 3) both HVDCs and generators have droop control.

As shown in Fig. 9, subcase 1 illustrates that there exists frequency stability problem if there is no any droop control. In subcase 2 and 3, the frequency of the receiving-end system remains stable owning to the droop based EDCPS strategy. And the system has a better performance if both HVDCs and generators have droop control. The active power of HVDCs in subcase 3 is shown in Fig.10, which also verify the effectiveness of the EDCPS strategy.

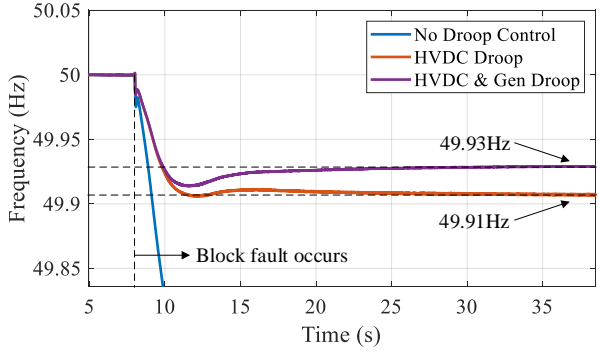

Figure 9. The frequency of the receiving-end system.

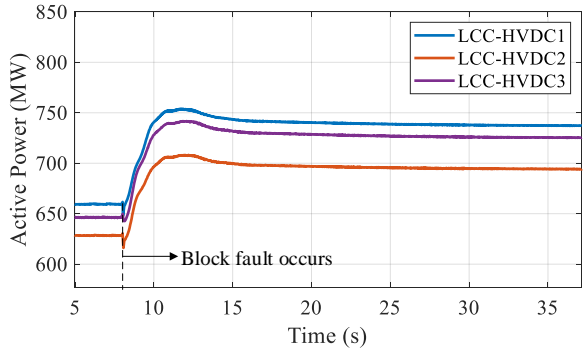

Figure 10. The active power of the LCC-HVDC system.

Further, the optimality of the droop coefficients is discussed. Set all the droop coefficients of LCC-HVDCs are fixed value, such as the average value of the optimal droop coefficients (28 p.u.), to contrast with the optimal droop coefficients. The frequency deviations of the receiving-end system and send-ending systems with fixed and optimal droop coefficients respectively are shown in Fig. 11.

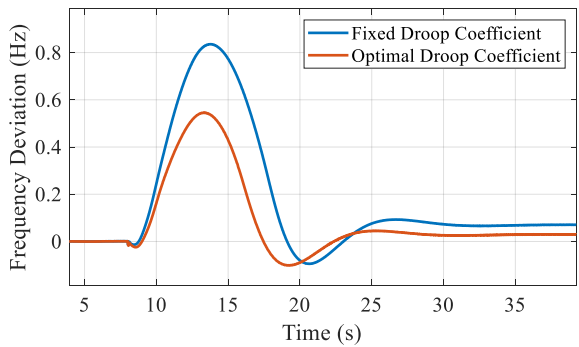

Figure 11. The frequency deviation between the receiving-end and sending-end system.

As shown in Fig. 11, there is smaller frequency deviation after the transient process if the droop coefficients are optimized in EDCPS strategy, which illustrates the correctness of the proposed coordinated optimization method.

V. CONCLUSION

In this paper, a novel coordinated droop control based EDCPS strategy for MIDC system is proposed. A P-f droop characteristic of LCC-HVDC system is designed, and a coordinated optimization method is proposed to determine the optimal droop coefficients of multiple HVDC lines. The correctness of the designed P-f droop characteristic and the effectiveness of the proposed EDCPS strategy are tested and validated through two scenarios on the CloudPSS platform.